\documentclass[aps,pre,twocolumn,superscriptaddress,showpacs,floatfix]{revtex4}
\usepackage{graphicx}
\usepackage{bm}
\begin{document}

\title{Non-��integrability and the Fourier heat conduction law}
\author{Shunda Chen}
\affiliation{CNISM and Center for Nonlinear and Complex Systems,
Universit\`a degli Studi dell'Insubria, via Valleggio 11, 22100 Como, Italy}
\affiliation{Istituto Nazionale di Fisica Nucleare, Sezione di Milano,
via Celoria 16, 20133 Milano, Italy}
\author{Jiao Wang}
\affiliation{Department of Physics and Institute of Theoretical Physics
and Astrophysics, Xiamen University, Xiamen 361005, Fujian, China}
\author{Giulio Casati}
\affiliation{CNISM and Center for Nonlinear and Complex Systems,
Universit\`a degli Studi dell'Insubria, via Valleggio 11, 22100 Como, Italy}
\affiliation{Istituto Nazionale di Fisica Nucleare, Sezione di Milano,
via Celoria 16, 20133 Milano, Italy}
\affiliation{International Institute of Physics, Federal University of
Rio Grande do Norte, Natal, Brasil}
\author{Giuliano Benenti}
\affiliation{CNISM and Center for Nonlinear and Complex Systems,
Universit\`a degli Studi dell'Insubria, via Valleggio 11, 22100 Como, Italy}
\affiliation{Istituto Nazionale di Fisica Nucleare, Sezione di Milano,
via Celoria 16, 20133 Milano, Italy}

\date{\today}

\begin{abstract}
We study in momentum��-conserving systems, how nonintegrable dynamics
may affect thermal transport properties. As illustrating examples, two
one��-dimensional (1D) diatomic chains, representing 1D fluids and lattices,
respectively, are numerically investigated. In both models, the two species
of atoms are assigned two different masses and are arranged alternatively.
The systems are nonintegrable unless the mass ratio is one. We find that
when the mass ratio is slightly different from one, the heat conductivity
may keep significantly unchanged over a certain range of the system size
and as the mass ratio tends to one, this range may expand rapidly. These
results establish a new connection between the macroscopic thermal transport
properties and the underlying dynamics.
\end{abstract}

%44.10.+i Heat conduction
%05.60.Cd classical Transport processes
%05.40.-a Statistical physics: Fluctuation phenomena
%51.20.+d Thermal conduction/Thermal diffusion in gases
\maketitle

\section{Introduction}

The Fourier heat conduction law is an empirical law that describes how the
heat current is sustained by the temperature gradient, i.e.,
\begin{equation}
j=-\kappa\nabla T,
\label{Four}
\end{equation}
where $j$ is the heat current, $\nabla T$ is the temperature gradient, and
$\kappa$ is known as the thermal conductivity, which is a finite constant
independent of the system size.

However, not all systems obey the Fourier law. It is known that the transport
properties are strongly affected by conservation laws~\cite{Mazur69, Zotos,
Ilievski, Benenti}. In the extreme case that a system is integrable, the heat
conductivity is a linear function of the system size. Even in the particular
case in which the total momentum is the only conserved quantity, the heat
conductivity may diverge as well. In particular, in one��-dimensional (1D)
and two��-dimensional (2D) cases, since 1970 when Alder and Wainwright reported
their findings~\cite{Alder70}, it has been realized that momentum conservation
may lead to slow decay of time correlations so that transport is not diffusive
and is characterized by diverging transport coefficients. For 1D
momentum-��conserving systems, the heat conductivity generally depends on the
system size $N$ in a power��-law manner: $\kappa\sim N^\alpha$. There is no
general consensus on the numerical value of  $\alpha$ and different theoretical
models predict that $\alpha$ is $1/2$ if the interparticle interaction is
symmetric and $1/3$ otherwise~\cite{Lee0508, DLLP0607, Beijeren}. It is worth
noting that these theoretical predictions equally apply to both fluids and
lattices. On the other hand, a recent numerical study~\cite{Zhong12} suggested
that when the interparticle interactions are asymmetric, there is a significant
difference between fluids and lattices. To summarize, for 1D systems, the heat
conduction properties are believed to depend on integrability,
momentum-��conservation, interaction symmetry, and the nature
of fluids or lattices.

For the particular case of 1D momentum��-conserving systems, which is the
subject of the present paper, all analytical and numerical results so far
available do not allow one to draw definite conclusions yet. This problem was
analyzed with various 1D models in a recent study~\cite{savin}, where
it was shown that the Fermi��-Pasta��-Ulam (FPU) chain with symmetric or
asymmetric potential exhibits anomalous heat transport, which is consistent
with other recent investigations~\cite{dhar, wang}. The plateau in the
system size dependence of the heat conductivity found in~\cite{chen} for
the FPU model with a certain set of parameters turns out to be a finite
size effect and, at larger $N$, the heat conductivity starts increasing
again. In particular in~\cite{wang} it was surmised that the value $1/3$
should be found asymptotically for very large system size, even though,
in fact, a value of the exponent $\alpha= 0.15$ was numerically found
(up to $N = 65536$). The results of~\cite{savin} also led to an exponent
$\alpha<1/3$ for the asymmetric FPU chain. In~\cite{dhar}, the value 1/3
was found for the same FPU model but in a different parameter range and
for high temperatures. In the same paper, the possibility of a finite
temperature phase transition was not ruled out. Finally, in~\cite{savin}
normal heat conductivity was reported for 1D momentum��-conserving systems
with the Lennard��-Jones, Morse, and Coulomb potential.

The overall picture is therefore far from being clear. \textit{Rebus sic
stantibus}, in order to gain a better understanding in such a complex
situation, it might be convenient to consider the 1D diatomic hard��-point
gas. Indeed, this is a clean and simple system of billiard type and, as such, it
should reflect general properties since billiards have been found fundamental in
understanding both classical and quantum dynamical systems. Moreover, an
important feature of billiard��-type systems is that their dynamical
properties do not depend on the temperature, which makes their analysis
even more simplified. By analyzing the hard��-point gas, we show that
close to the integrable, equal masses limit, the system exhibits normal
heat conduction over longer and longer sizes as the integrable limit is
approached. Asymptotically, however, the power law divergence of the
thermal conductivity sets in with the power 1/3. To be more precise, we
cannot exclude the possibility of a phase transition as the mass ratio
is increased; however, our numerical evidence suggests that this possibility
should be quite unlikely. The analysis of the diatomic Toda lattice confirm
these conclusions. These results lead us to speculate that as one approaches
the integrable limit, anomalous behavior is perhaps more general than so far
expected~\cite{Zhong12,chen, savin} even though it might be hard to detect
in numerical simulations.

\section{1D diatomic gas model}

After being initially proposed in 1986~\cite{Casati86}, the 1D diatomic gas
model has attracted increasing interest for investigating various aspects of
1D transport. The model consists of $N$ hard��-core point particles in
one dimension with alternative mass $M$ and $m$ (for odd��- and
even��-numbered  particles, respectively). We fix the averaged particle
number density to be unity so that $N$ refers to the length of the system as
well. In order to measure the heat conductivity, two statistical thermal baths
with different temperatures $T_L$ and $T_R$ are put into contact with the left
and the right end of the system. When the first (last) particle collides with
the left (right) side of the system, it is injected back with a new speed $|v|$
determined by the distribution~\cite{heatbath}
\begin{equation}
P_{L,R}(v) = \frac{|v|\mu_{1,N}}{k_B T_{L,R}}\exp\left(
- \frac{v^{2} \mu_{1,N}}{2 k_B T_{L,R}}\right).
\end{equation}
Here $\mu_{1}$ and $\mu_{N}$ are the masses of the first and the last
particle and $k_B$ is the Boltzmann constant which is set to be unity
throughout.

In our simulations, each particle is given initially a random position
uniformly distributed and a random velocity according to the Boltzmann
distribution with temperature $T(x_i)=T_L+x_i (T_R-T_L)/N$ ($x_i$ is
the position of the $i$th particle). Then the system is evolved by using
an effective event-��driven algorithm~\cite{Casati03}. After the system
reaches the steady state, we compute the steady heat flux $j$ that crosses
the system; i.e., the averaged energy exchanged in the unit time between
a boundary particle and the heat bath, or that between any two neighboring
particles. The heat conductivity is then measured, by assuming the Fourier
law, as $\kappa \approx jN/(T_L-T_R)$. We set $T_L=6$ and $T_R=4$ so that
the nominal temperature of the system is $T=5$. The heat conductivity at
any other temperature $T'$ can be obtained through the scaling relation
$\kappa (T') = \kappa(T) \sqrt{T'/T}$. We will focus on how the heat
conductivity $\kappa$ depends on the system size $N$ and on the mass ratio
$M/m$ (hereafter we set $m\equiv 1$). We emphasize that in our simulations,
long enough integration times ($>10^8$) have been taken so that the
relative errors of all the measured values of $\kappa$ are less than
$1\%$.

Now let us turn to the simulations results. First of all, if the mass
ratio is unity then the system is integrable and, with the heat bath
given by Eq. (2), the heat conductivity writes:
\begin{equation}
\kappa_{\text {int}}=N\sqrt{\frac{2k^3_B}{m\pi}}/\left(\frac{1}
{\sqrt{T_L}}+\frac{1}{\sqrt{T_R}}\right).
\end{equation}
In Fig. 1(a) this result is compared with our simulations and the
agreement is perfect. This can be considered as a numerical test.
Now, we change the mass ratio to make it slightly larger than one
[see Fig. 1(a)]; it can be seen that for small $N$ ($<10^2$), $\kappa$
follows its integrable limit case, but as $N$ is increased further,
$\kappa$ tends to saturate and becomes constant for $N>10^4$. This
could be taken as an empirical demonstration that at least for these
mass ratios and for large enough system size, heat conduction is
governed by the Fourier law, which is in clear contrast with existing
theoretical and numerical predictions. (See for example
Refs.~\cite{DLLP0607, Beijeren}).

%%%%%%%%%%%%%%%%%%%%%%%%%%%%%%%%%%%%%%%%%%%%%%%%%%%%%%%%%%%%%%%%%%%%%%%%%%%%%%%Fig1
\begin{figure}
\vskip-.2cm\includegraphics{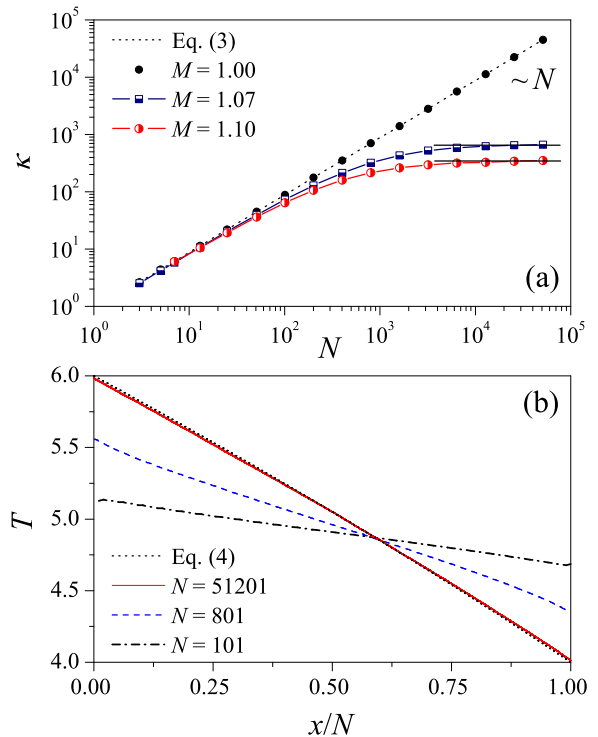} \vskip-.4cm
\caption{(Color online) (a) The heat conductivity $\kappa$ as a function
of the system size $N$ in the 1D diatomic gas model. The two horizontal
lines denote the saturation value of $\kappa_{GK}(N)$ [Eq.~(\ref{eq:GK})]
at large $N$, for mass ratios $M=1.07$ and $1.1$. (b) Comparison between
the numerically computed temperature profile and the analytic expression
[see Eq. (4)] for $M=1.07$ at different system sizes.}
\end{figure}
%%%%%%%%%%%%%%%%%%%%%%%%%%%%%%%%%%%%%%%%%%%%%%%%%%%%%%%%%%%%%%%%%%%%%%%%%%%%%%%Fig1

The validity of the Fourier law also determines the internal temperature
profile of the steady state. Indeed by assuming the Fourier law and equating
the averaged local heat flux along the system, one obtains~\cite{ADhar01}
\begin{equation}
T(x)=\left[T_L^{3/2}\left(1-\frac{x}{N}\right)
+T_R^{3/2}\frac{x}{N}\right]^{2/3}.
\end{equation}
In Fig. 1(b), this prediction is compared with our simulations results for
$M=1.07$. Numerically, the temperature of the $i$th particle is measured as
the time average of its kinetic energy, i.e., $T(x_i)=\langle \mu_i v_i^2/
k_B \rangle$, with $\mu_i\in\{M,m\}$ and $v_i$ being its mass and velocity,
respectively. It is seen that numerical results are in very good agreement,
for $N>10^4$, with this theoretical prediction.

%%%%%%%%%%%%%%%%%%%%%%%%%%%%%%%%%%%%%%%%%%%%%%%%%%%%%%%%%%%%%%%%%%%%%%%%%%%%%%%Fig2
\begin{figure}
\vskip-.2cm \includegraphics{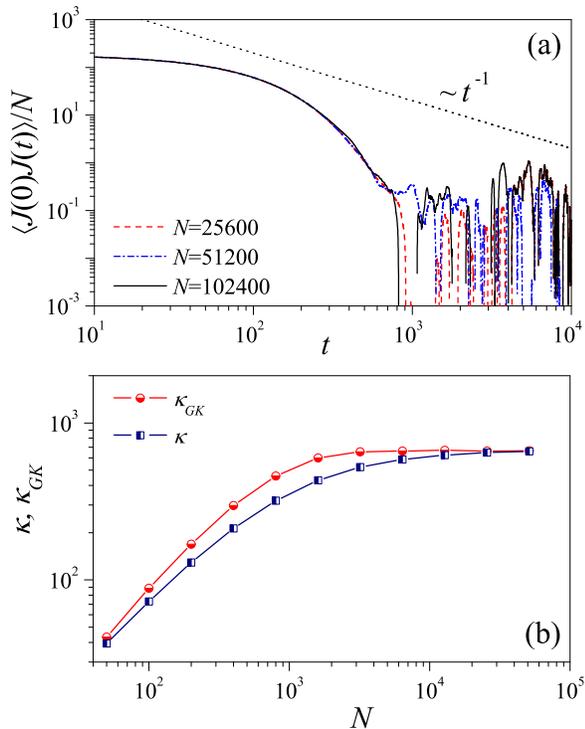} \vskip-.4cm
\caption{(Color online) (a) Correlation functions of the total heat current
for the 1D diatomic gas model. The dotted line indicates the scaling $\sim
t^{-1}$: A faster decay of the correlation function implies convergence of
the heat conductivity in the thermodynamic limit. (b) The comparison of the
heat conductivity obtained by using the Green��-Kubo formula [Eq. (5)] and
by using the nonequilibrium setting. In both panels $M=1.07$.}
\end{figure}
%%%%%%%%%%%%%%%%%%%%%%%%%%%%%%%%%%%%%%%%%%%%%%%%%%%%%%%%%%%%%%%%%%%%%%%%%%%%%%%Fig2

We now turn to the linear response theory to check if this approach leads
to consistent results thus confirming the validity of the Fourier law for
large $N$. Based on the Green��-Kubo formula, which relates transport
coefficients to the current time��-correlation functions, the heat
conductivity of a 1D finite system can be expressed as~\cite{Lepri03,
Prosen05}
\begin{equation}
\kappa_{GK}(N)=\frac{1}{k_B T^2 N}\int_0^{\tau_{tr}}dt\langle J(0)J(t)
\rangle.
\label{eq:GK}
\end{equation}
In this formula, $J\equiv\sum_i\mu_i v_i^3/2$ represents the total
heat current and $\langle J(0)J(t)\rangle$ is its correlation function
measured in the equilibrium state with the periodic boundary condition.
The integration is truncated at time $\tau_{tr}$ which is suggested to
assume the value of $\tau_{tr}=N/(2v_s)$ ($v_s$ is the sound speed of the
system)~\cite{Chen14}. To numerically compute $\kappa_{GK}(N)$, we consider
isolated systems with periodic boundary conditions. The initial condition
is randomly assigned with the constraints that the total momentum is zero
and the total energy corresponds to $T=5$. The system is then evolved and
after the equilibrium state is reached, we compute $\langle J(0)J(t)\rangle$
and the integral in Eq. (5).

%%%%%%%%%%%%%%%%%%%%%%%%%%%%%%%%%%%%%%%%%%%%%%%%%%%%%%%%%%%%%%%%%%%%%%%%%%%%%%%Fig3
\begin{figure}
\vskip-.2cm\includegraphics[width=1.01\columnwidth,clip]
{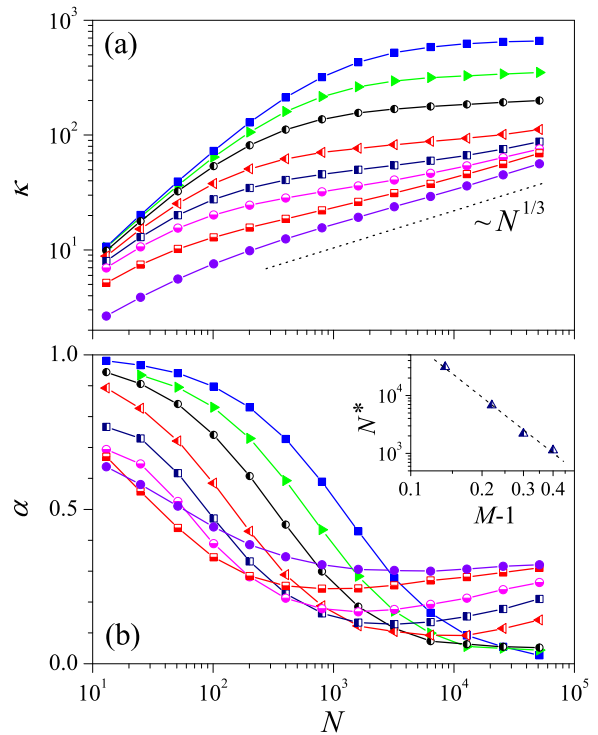} \vskip-.4cm
\caption{(Color online) (a) The heat conductivity $\kappa$ versus the
system size $N$ for the 1D diatomic gas model. From top to bottom, the
mass ratio $M$ is respectively $1.07$, $1.10$, $1.14$, $1.22$, $1.30$,
$1.40$, the golden mean ($\approx1.618$), and $3$. The corresponding
tangent $\alpha$ of the $\kappa$��-$N$ curve is given in (b) with the
same symbols. In the inset we plot the turning point $N^\ast$, after
which $\alpha$ starts growing with $N$, as a function of $M-1$. The
best fitting (the dotted line) suggests $N^\ast = 54/(M-1)^{3.2}$.}
\end{figure}
%%%%%%%%%%%%%%%%%%%%%%%%%%%%%%%%%%%%%%%%%%%%%%%%%%%%%%%%%%%%%%%%%%%%%%%%%%%%%%%Fig3

The results for $M=1.07$ are presented in Fig. 2. It can be seen from
Fig. 2(a) that for a large system ($N>10^4$), the correlation function
changes slowly at short times ($t<10^2$), which reflects the fact that
the system still mimics its integrable limit; however, from $t\sim 10^2$
to $10^3$, the correlation function undergoes a rapid decay and eventually,
when $t> 10^3$, it begins to oscillate around zero. (The negative values
of $\langle J(0)J(t)\rangle$ are not shown in this log��-log scale.) In
Fig. 2(b), the dependence of $\kappa_{GK}$ on the system size is shown.
It can be seen that $\kappa_{GK}$ agrees with $\kappa$ despite some
deviations at small $N$.

Next we consider the dependence on the mass ratio. By using the same
nonequilibrium setting we have extensively investigated the system size
dependence of $\kappa$  for the mass ratio ranging from 1.07 to 64. The
results for $1.07\le M\le 3$ are shown in Fig. 3(a). A three-��stage
process can be recognized : For small system sizes $\kappa \sim N$, similar
to the integrable case. For large system sizes, $\kappa$ shows a tendency to
$\sim N^{1/3}$. In between these two regimes, there appears an intermediate,
bridging regime, where $\kappa$ changes at a lower rate (see particularly
the cases of $M=1.22$ and $1.30$). Actually, in this intermediate
regime, as $M$ is decreased, the conductivity $\kappa$ tends to be
constant over a larger and larger interval. For $M\ge3$ instead (data
not shown here) the dependence $\kappa\sim N^{1/3}$ appears more and more
clearly in agreement with the existing theories~\cite{DLLP0607, Beijeren}.
%However, for $M<3$ but $M\ge 1.3$, all the three above discussed regimes
%can be clearly seen. Finally, for $M< 1.3$, as we have seen in Fig. 1(a),
%$\kappa$ experiences a transition from $\sim N$ to $\sim N^0$.

In order to better understand the dependence of $\kappa$ on
$N$, along each curve provided in Fig. 3(a) we computed its tangent
$\alpha(N)$ and plot the results in Fig. 3(b). Note that $\alpha(N)$
exhibits a non��-monotonic behavior and reaches a minimum at a certain
system size $N^\ast$. Interestingly enough, the value of $N^\ast$ appear
to grow very fast with decreasing $M$ [see the inset in Fig. 3(b)]. This
result shows that a very small tangent $\alpha$, i.e., a Fourier��-like
behavior of thermal conduction, can be observed over an increasingly large
system size when the integrable limit is approached. At the same time, for
$N>N^\ast$, anomalous behavior emerges gradually.
%Another interesting observation is
%that as $M$ varies, $\kappa(N_\ast)=0.91N_\ast^{0.52}$ with the exponent
%being in between the two theoretical limits of $1$ and $1/3$.

The conclusion is that for any mass ratio different from unity the behavior
$\kappa \sim N^{1/3}$ seems to always take place even though it cannot be
detected numerically when the mass ratio approaches unity since in this
limit $N^\ast$ becomes exceedingly large. On the other hand, based on our
available data, the possibility that there is a phase transition around
$M\approx 1.3$  can not be ruled out with certainty.

%%%%%%%%%%%%%%%%%%%%%%%%%%%%%%%%%%%%%%%%%%%%%%%%%%%%%%%%%%%%%%%%%%%%%%%%%%%%%%%Fig4
\begin{figure}
\vskip-.2cm \includegraphics{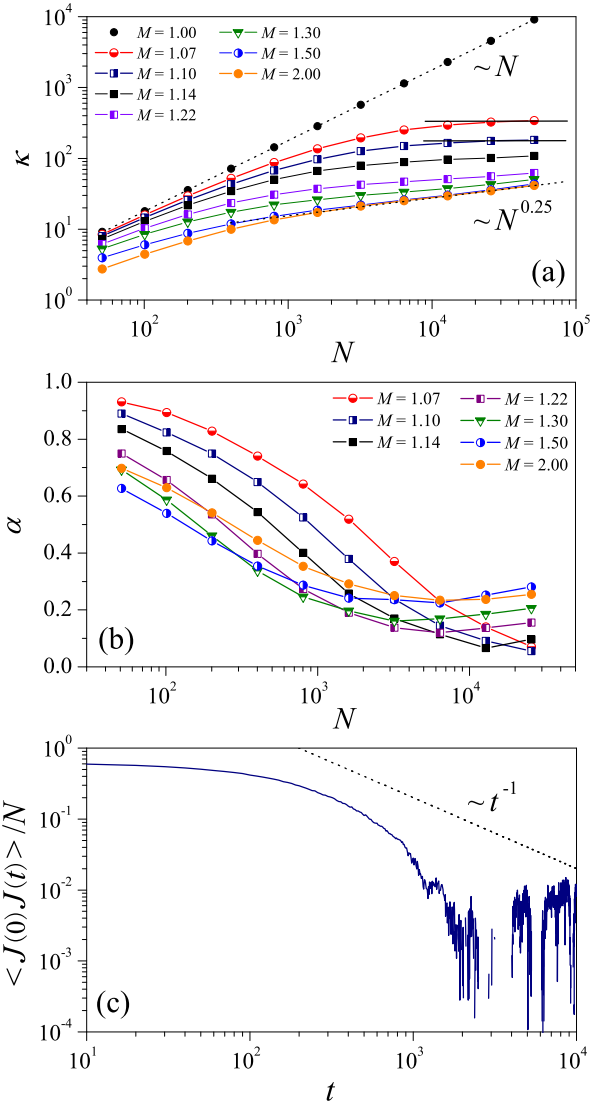}\vskip-.5cm
\caption{(Color online) (a) The heat conductivity measured in the
nonequilibrium setting for the 1D diatomic Toda lattice with mass $M=1$
(the integrable case), $1.07$, $1.10$, $1.14$, $1.22$, $1.30$, $1.50$, and $2$.
The dotted lines indicate,
respectively, the ballistic behavior $\kappa\sim N$ and the power law best
fitting to the case of $M=2$, $\kappa\sim N^\alpha$, with $\alpha=0.25$.
The horizontal lines denote the saturated values of $\kappa_{GK}(N)$ for
$M=1.07$ and $1.10$. (b) The corresponding tangent $\alpha$ of the
$\kappa$��-$N$ curve with the same symbols. (c) The heat current
correlation function for $M=1.10$ with $N=25600$, showing a decay faster
than $\sim 1/t$.}
\end{figure}
%%%%%%%%%%%%%%%%%%%%%%%%%%%%%%%%%%%%%%%%%%%%%%%%%%%%%%%%%%%%%%%%%%%%%%%%%%%%%%%Fig4

\section{1D diatomic Toda chain}

The above described scenario in which the Fourier law appears in the
"vicinity" of the integrable limit is not exclusive of the gas model. In
the following we show that it is also the case for lattices. The model we
consider here is a diatomic variant of the Toda lattice~\cite{Lepri03,
Hatano98} with the Hamiltonian
\begin{equation}
H=\sum_{i}\left[\frac{p_{i}^2}{2\mu_i}+U(x_{i}-x_{i-1})\right],
\end{equation}
where the potential is $U(x)=\exp(-x)+x$, and the particles take masses
$M$ and $m\equiv 1$ alternatively. As for the gas model, this system is
integrable when the mass ratio is one. We measure the heat conductivity
in both the nonequilibrium and equilibrium settings again, and find that
the results turn out to agree with each other. In the nonequilibrium
simulations, we couple the system to two Langevin heat baths~\cite{Dharrev}
with the temperature $T_L=1.2$ and $T_R=0.8$. The heat current is defined
as $j\equiv\langle j_i\rangle$ with $j_i\equiv v_i{\partial}U(x_{i+1}-x_i)
/{\partial x_i}$~\cite{TMai07}. In Fig. 4(a) the measured $\kappa$ for
different values of $M$ is given. Again, for mass ratios close to unity,
$\kappa$ is close to the integrable case when the system is small ($N<10^2$)
but tends to a value which agrees with that obtained by using the
Green��-Kubo formula for the large system's size ($N>10^4$). For larger
mass ratio (see the case of $M=2$) the heat conductivity is anomalous.
Similarly to the hard-point gas model, the tangent $\alpha(N)$ exhibits
a nonmonotonic behavior, with the minimum reached at a system size $N^*$
rapidly growing when the integrable limit $M=1$ is approached [see Fig. 4(b)].
With regard to the equilibrium simulations, we assume periodic boundary
conditions, null total momentum and total energy corresponding to $T=1$.
The total heat current is $J=\sum_i j_i$ and its correlation function for
$M=1.1$ is shown in Fig. 4(c), where it exhibits a faster than $\sim 1/t$
decay as expected in the case of normal heat conduction. The overall
emerging picture is the same as presented above for the gas model. This
similarity is unlikely a coincidence due to the contrasting difference
in the dynamics of the two systems; rather, it strongly suggests some
general mechanisms in the heat conduction properties as one departs
from the integrable limit.

\section{Summary and discussions}

We have shown that in two 1D
momentum-��conserving paradigmatic systems, the heat conductivity can
be independent of the system size over a considerably wide range. Such a
Fourier��-like  behavior appears as a quite general feature for lattice or
gas models close to the integrable limit. Apart from theoretical implications
in transport theory, our finding may have experimental relevance as well,
because the system size over which the heat conductivity keeps constant,
grows very fast as the system approaches its integrable limit.
%In this regard, the effect
%of the $\sim t^{-2/3}$ tail in the correlation function of the heat current
%can be neglected in some circumstances for practical purposes~\cite{Chen13}.
%Quite curiously, in order to observe the anomalous behavior for reasonably
%small system sizes one should consider parameters sufficiently far from the
%integrable limit, while to observe a Fourier��-like behavior, i.e. a typical
%behavior, one should go close to integrability, but integrable systems are
%atypical.

Our present understanding of the heat conduction problem is mainly based
on numerical empirical evidence while rigorous analytical results are hard
to obtain. Numerical analysis consists of steady-state, nonequilibrium
simulations or of equilibrium simulations based on linear response theory
and the Green��-Kubo formula. If both methods give reasonable evidence for
the Fourier law and if, moreover, they lead to the same numerical value of
the heat conductivity $\kappa$, then this has been generally considered as
a conclusive evidence that the Fourier law is valid. This conclusion, however,
could not be correct. As we have shown in this paper, the agreement between
equilibrium and nonequilibrium simulations does not allow, \emph{per se},
to draw any definite conclusion. Indeed this agreement might be a finite
size effect and the Fourier law may appear to hold up to some system size
$N$ after which anomalous behavior sets in. The main point is that we have
no indications at all about the critical value of $N$ after which conductivity
becomes anomalous. What we know from the numerical analysis of this paper
is that this critical value seems to diverge rapidly as one approaches the
integrable limit. This result is quite surprising to us and it is a feature
which we do not understand yet. While it is natural to expect an initial
ballistic behavior for larger and larger system sizes as one approaches
the integrable limit, it is absolutely not clear why the value of $\kappa$
appears to saturate to a constant value and why this Fourier��-like behavior
may persist in an increasingly wide range of the system size before entering
the anomalous regime.

\section*{Acknowledgements}

Useful discussions with Stefano Lepri are gratefully acknowledged. This work
is supported by NSFC (Grants No. 11275159 and No. 11335006) and by MIUR-PRIN.


\begin{thebibliography}{99}


\bibitem{Mazur69}   P. Mazur, Physica (Amsterdam) {\bf 43}, 533 (1969);
                    M. Suzuki, {\it ibid}, {\bf 51}, 277 (1971).
\bibitem{Zotos}     X. Zotos, F. Naef, and P. Prelov\v sek,
                    Phys. Rev. B {\bf 55}, 11029 (1997).
\bibitem{Ilievski}  E. Ilievski and T. Prosen,
                    Commun. Math. Phys. {\bf 318}, 809 (2013).
\bibitem{Benenti}   G. Benenti, G. Casati, and J. Wang,
                    Phys. Rev. Lett. \textbf{110}, 070604 (2013).
\bibitem{Alder70}   B. J. Alder and T. E. Wainwright,
                    Phys. Rev. A \textbf{1}, 18 (1970).
\bibitem{Lee0508}   G. R. Lee-Dadswell, B. G. Nickel, and C. G. Gray,
                    Phys. Rev. E {\bf 72}, 031202 (2005); J. Stat. Phys. {\bf 132}, 1 (2008)
\bibitem{DLLP0607}  L. Delfini, S. Lepri, R. Livi, and A. Politi,
                    Phys. Rev. E  {\bf 73}, 060201(R) (2006); J. Stat. Mech. P02007 (2007).
\bibitem{Beijeren}  H. van Beijeren,
                    Phys. Rev. Lett. \textbf{108}, 180601 (2012).
\bibitem{Zhong12}   Y. Zhong, Y. Zhang, J. Wang, and H. Zhao,
                    Phys. Rev. E \textbf{85}, 060102(R) (2012).
\bibitem{savin}     A. V. Savin and Y. A. Kosevich,
                    Phys. Rev. E {\bf 89}, 032102 (2014).
\bibitem{dhar}      S. G. Das, A. Dhar, and O. Narayan,
                    J. Stat. Phys. {\bf 154}, 204 (2013).
\bibitem{wang}      L. Wang, B. Hu, and B. Li,
                    Phys. Rev. E {\bf 88}, 052112 (2013).
\bibitem{chen}      S. Chen, Y. Zhang, J. Wang, and H. Zhao,
                    arXiv: 1204.5933.
\bibitem{Casati86}  G. Casati,
                    Found. Phys. \textbf{16}, 51 (1986).
\bibitem{heatbath}  J. L. Lebowitz and H. Spohn,
                    J. Stat. Phys. \textbf{19}, 633 (1978);
                    R. Tehver, F. Toigo, J. Koplik, and J. R. Banavar,
                    Phys. Rev. E \textbf{57}, R17 (1998).
\bibitem{Casati03}  G. Casati and T. Prosen,
                    Phys. Rev. E \textbf{67}, 015203 (2003).
\bibitem{ADhar01}   A. Dhar,
                    Phys. Rev. Lett. \textbf{86}, 3554 (2001).
\bibitem{Lepri03}   S. Lepri, R. Livi, and A. Politi,
                    Phys. Rep. \textbf{377}, 1 (2003).
\bibitem{Prosen05}  T. Prosen and D. K. Campbell,
                    Chaos \textbf{15}, 015117 (2005).
\bibitem{Chen14}    S. Chen, Y. Zhang, J. Wang, and H. Zhao,
                    Phys. Rev. E \textbf{89}, 022111 (2014).
\bibitem{Hatano98}  T. Hatano,
                    Phys. Rev. E \textbf{59}, 1(R) (1999).
\bibitem{Dharrev}   A. Dhar, Adv. Phys. \textbf{57}, 457 (2008).
\bibitem{TMai07}    T. Mai, A. Dhar, and O. Narayan,
                    Phys. Rev. Lett. \textbf{98}, 184301 (2007).
\end{thebibliography}
\end{document}